\documentstyle[12pt]{article}

\begin{document}

\begin{titlepage}

\vskip 1.4truecm

\begin{center}
{\large \bf COHOMOLOGICAL PARTITION FUNCTIONS   \\ \vskip 0.4cm
             FOR A CLASS OF BOSONIC THEORIES }
\end{center}
  \par \vskip .1in \noindent

\vskip 2.0cm

\begin{center}
{\bf Antti J. Niemi$^{* ~\#}$ and O. Tirkkonen } \\
\vskip 0.3cm
{\it Research Institute for Theoretical Physics,
University of Helsinki \\
Siltavuorenpenger 20 C, SF-00170 Helsinki, Finland}
\end{center}

\vskip 4.5cm

We argue that for a general class of nontrivial bosonic theories the
path integral can be related to an equivariant generalization of
conventional characteristic classes.

\vfill

\begin{flushleft}
\rule{5.1 in}{.007 in}\\
$^{*}$ {\small E-mail: ANIEMI@PHCU.HELSINKI.FI \\ } \vskip 0.2cm
${\#}$ {\small Permanent address after 1.7.1992: Institute for
Theoretical Physics, Uppsala University, Uppsala, Sweden}
\end{flushleft}

\end{titlepage}

During the last twenty years, characteristic classes [1] have been involved in
several major developments both in physics and
mathematics. The most familiar example of characteristic classes
is the four dimensional $F \tilde F$ and its generalizations to other even
dimensions, together with their odd-dimensional descendants such as the
Chern-Simons action. Another widely encountered example is the
gravitational $R\tilde R$ and its relatives. These characteristic classes
are relevant to a large number of phenomena such as anomalies, fermion
number fractionization, index theorems, CP-violation, fractional spin and
statistics, high-temperature superconductivity and so on. More recently,
the quantization of these characteristic classes has revealed
several important results and provides us with new insight
to problems from knot theory to quantum gravity [2].

In this Letter we wish to draw attention to a natural generalization of
characteristic classes. This generalization has been discussed quite
extensively in the contemporary mathematics literature, but until now
it has not emerged in the physics literature. This generalization arises when
ordinary cohomology is extended to equivariant cohomology [3-5]. In
addition to the exterior derivative, equivariant cohomology also includes
the action of a preferred vector field. In applications to physics such
a preferred vector field is naturally
present. It is the vector field that determines the hamiltonian flow. As
a consequence equivariant cohomology provides a natural
framework for extending conventional cohomology and its characteristic
classes in a dynamically nontrivial fashion:  Indeed, equivariant
characteristic classes [4-5] share all the beauty, simplicity and
power of their conventional counterparts,  but have the additional
feature that they account for a nontrivial hamiltonian action. Consequently
we expect, that equivariant characteristic classes will become relevant
to a number of important problems in physics.

Conventional characteristic classes are usually associated to the
quantization of fermions: Anomalies are an important
ingredient of the path integral measure in fermionic quantum field
theories, and the Atiyah-Singer index theorem can also be derived from a
supersymmetric path integral. In the present Letter we shall argue that
equivariant characteristic classes can be associated to the quantization of
{\it bosons}. Indeed, we shall identify a general class of bosonic
hamiltonians such that the corresponding path integral evaluates to
an equivariant characteristic class. The condition that we impose
is quite general: Locally in a neighborhood, our condition is satisfied
by a {\it generic} hamiltonian. Thus we expect that equivariant
characteristic classes are an important ingredient also in the general
quantization procedure, and corrections to our result seem
to be related to the global aspects of the theory.

We shall first shortly explain aspects of equivariant cohomology.
A more extensive discussion can be found {\it e.g.} in [5], and
some applications to physics can be found {\it e.g.} in [6,7]. We
shall begin from ordinary cohomology:

In ordinary cohomology we are interested in the action of a nilpotent exterior
derivative $d$  in the space of exterior forms which are associated
to some manifold. In this space, $d$ determines a linear mapping
which sends an $n$-form onto an $n+1$-form. The $n$-forms which are
annihilated by $d$ are closed if they can not be represented as an
exterior derivative of  an $n-1$-form. Otherwise, they are exact.
The space of closed $n$-forms modulo exact $n$-forms determines
the $n$th cohomology class of the manifold. Conventional
characteristic classes are examples of nontrivial cohomology classes.
In the physics literature, the most familiar is the
Chern character of a gauge theory,
$$
Ch(F) ~=~ Tr \{ e^{ { 1 \over 4 \pi} F} \}
\eqno (1)
$$
where $F$ is the Yang-Mills curvature two-form in some (properly
normalized) representation of the gauge group. Its gravitational analog
is the $\hat A$-genus
$$
{\hat A} (R) ~=~ \prod\limits_{i=1}^{\frac{n}{2}} { \frac{1}{2}x_{i} \over
sinh( \frac{1}{2} x_{i} ) }
\eqno (2)
$$
where the $x_{i}$ are the skew-eigenvalues of the curvature two-form matrix
$R_{ab}$. The expansion of
these  characteristic classes in their arguments yields closed forms, and the
integrals of these closed forms are topological invariants of the manifold.

Equivariant cohomology is an extension of conventional cohomology.
In addition of  the exterior derivative, now there is also
an interior multiplication (contraction) along a preferred vector field.
With ${\cal X}^{a}$ components of the
vector field, we denote the interior multiplication by
$$
i_{\cal X} ~=~ {\cal X}^{a} i_{a}
\eqno (3)
$$
where $i_{a}$ is interior multiplication dual to some basis of one-forms
$d\phi^{a}$ such that $i_{a} d\phi^{b} = \delta_{a}^{b}$. Interior
multiplication maps an $n$-form onto an $n-1$-form, and is nilpotent: $i_{\cal
X}^{2}=0$. Combining exterior derivative with interior multiplication we then
obtain the equivariant exterior derivative
$$
d_{\cal X} ~=~ d + i_{\cal X}
\eqno (4)
$$
Since $d$ maps an $n$-form onto an $n+1$-form and $i_{\cal X}$ maps an $n$-form
onto an $n-1$-form, $d_{\cal X}$ does not preserve the form degree but maps
an even form onto an odd form, and vice versa. If we identify even
forms with bosons and odd forms with fermions, we
can then view $d_{\cal X}$ as a
supersymmetry operator. The corresponding supersymmetry algebra
closes to the Lie
derivative along the vector field ${\cal X}^{a}$,
$$
d_{\cal X}^{2} ~=~ d i_{\cal X} + i_{\cal X} d ~=~ {\cal L}_{\cal X}
\eqno (5)
$$
and it has been established [7] that the supersymmetry algebra
in a {\it generic} supersymmetric theory is of
the form (5) in the (super)loop space of the
theory, provided one introduces an
appropriate auxiliary field formalism.

The equivariant exterior derivative (4) determines an equivariant
generalization of cohomology on the manifold. One can show that this
equivariant cohomology can be identified with conventional cohomology on the
subspace of exterior forms which are annihilated by the Lie
derivative (5). Indeed,
in this subspace the equivariant exterior derivative (4) becomes nilpotent, and
consequently it determines a conventional exterior derivative in this subspace.
Representatives of nontrivial cohomology classes can then be constructed using
the pertinent conventional characteristic classes.
However, since the equivariant
exterior derivative involves interior multiplication along
the vector field ${\cal
X}^{a}$, these characteristic
classes necessarily contain this vector field in a
definite, but nontrivial fashion. Furthermore, since equivariant cohomology
coincides with ordinary cohomology in the subspace
where (5) vanishes, it is obvious
that in spite of their nontrivial dynamical content, equivariant characteristic
classes share the geometrical properties of their conventional counterparts and
can be similarly analysed using powerful geometrical techniques [5].

The purpose of this Letter is to draw attention to equivariant
characteristic classes as potentially important quantities in physics.
In order to exemplify this, we shall now proceed to a path integral derivation
of the equivariant generalizations of (1) and (2). We shall find that {\it for
a
very general class of bosonic hamiltonians, the path integral evaluates to a
combination of the equivariant Chern character and
the equivariant $\hat A$-genus},
{\it i.e.} an equivariant generalization of the
Atiyah-Singer index density for a
Dirac operator. For this, we consider a phase space with a generic coordinate
system $\phi^{a}$. Here the index $a$ may be either finite
or infinite, and in a quantum
field theory application it denotes both internal degrees of freedom and space
cordinates. The symplectic two-form on this phase space is
$$
\omega ~=~ \frac{1}{2} \omega_{ab} d\phi^{a} \wedge d\phi^{b}
\eqno (6)
$$
and the inverse matrix $\omega^{ab}$ determines Poisson
brackets by
$$
\{ A(\phi) , B(\phi) \} ~=~ \omega^{ab} \partial_{a} A  \partial_{b} B
\eqno (7)
$$
Since  $d\omega=0$, we can locally represent it
as an exterior derivative of the symplectic one-form $\theta ~=~ \theta_{a}
d\phi^{a}$,
$$
\omega ~=~ d\theta
\eqno (8)
$$
We are interested in the equivariant cohomology determined
by the canonical action
of a given hamiltonian $H$. For this, we introduce
the corresponding hamiltonian
vector field ${\cal X}_{H}$ which is defined by
$$
{\cal X}_{H}^{a} ~=~ \omega^{ab} \partial_{b} H
\eqno (9)
$$
We can then introduce the corresponding equivariant exterior derivative
$$
d_{H} ~=~ d + i_{H}
\eqno (10)
$$
and the Lie-derivative along (9)
$$
{\cal L}_{H} ~=~ d i_{H} + i_{H} d
\eqno (11)
$$
In the following we shall {\it assume} that the phase space admits a
metric tensor $g_{ab}$ which is Lie-derived by ${\cal
X}_{H}$,
$$
{\cal L}_{H} g ~=~ 0
\eqno (12)
$$
Locally  we can always find this metric tensor in a small enough phase space
neighborhood (and outside of the critical points of $H$)
but there might be global
difficulties associated in defining $g_{ab}$. Consequently (12) determines a
restriction on the class of hamiltonians that we shall consider. However, since
(12) can be satisfied by a large number of interesting hamiltonians defined
{\it e.g.} on phase spaces with a K\"ahler structure, we do not view (12) as an
overly strong restriction.

We are interested in evaluating the phase space path integral
$$
Z ~=~ \int [d\phi^{a}] \sqrt{ Det ( \omega_{ab} ) } exp \{ i
\int\limits_{0}^{T}
\theta_{a}{\dot \phi}^{a} - H \}
\eqno (13)
$$
with periodic boundary conditions. For this, we first introduce anticommuting
variables $c^{a}$ and write (13) as
$$
Z ~=~ \int [d\phi^{a}] [dc^{a}] exp
\{ i \int\limits_{0}^{T} \theta_{a} \dot\phi^{a}
- H + \frac{1}{2}c^{a}\omega_{ab}c^{b} \}
\eqno (14)
$$
In [6] we have investigated path integrals of the form (14).
There, we interpret (14) as a loop space integral, with the loop space
parametrized by the time evolution $\phi^{a} \to \phi^{a}(t)$. The
bosonic part of
the action then determines a loop space hamiltonian vector field,
$$
{\cal X}_{S}^{a} ~=~ \dot \phi^{a} - \omega^{ab} \partial_{b} H
\eqno (15)
$$
and we can identify $c^{a} (t)$ as the basis of one-forms on this loop
space. Defining  the loop space exterior derivative (in the following time
integrals are implicit)
$$
d ~=~ c^{a} \partial_{a}
\eqno (16)
$$
we then obtain the loop space version of the equivariant exterior
derivative,
$$
d_{S} ~=~ c^{a} \partial_{a} + {\cal X}_{S}^{a} i_{a}
\eqno (17)
$$
Here $i_{a}$ is a basis for loop space interior multiplication,
defined by $i_{a}c^{b}=\delta_{a}^{b}$.

In analogy with (10), (11) we conclude that the equivariant exterior
derivative (17) determines a loop space supersymmetry
transformation. The path integral (14) is (formally) invariant under this
supersymmetry transformation. From this we conclude [6] that if $\psi$ is a
loop
space one-form such that
$$
{\cal L}_{S} \psi ~=~ (d i_{S} + i_{S} d) \psi ~=~ 0
\eqno (18)
$$
then the path integral
$$
Z_{\psi} ~=~ \int [d\phi^{a}] [dc^{a}] exp\{ i \int\limits_{0}^{T} \theta_{a}
\dot\phi^{a} - H + \frac{1}{2}c^{a}\omega_{ab}c^{b} + d_{S} \psi \}
\eqno (19)
$$
is (formally) independent of $\psi$, and coincides
with the original path integral
(14). In particular, if we select $\psi$ properly we obtain
the path integral generalization [6] of the Duistermaat-Heckman integration
formula [5,8], which states that the path
integral (19) localizes to the critical
points of the bosonic action. Alternative choices of $\psi$ then yield
alternative integration formulas corresponding to localications to
different field configurations. If more than one integration
formula is valid for a
given theory, their answers must coincide. However, for some theories
it may happen
that some integration formulas are not valid. Then, different integration
formulas can yield different results and we must select the
one which gives the correct
result, if it exists.

In the present Letter we shall employ the condition (12) to introduce
a class of functionals $\psi$ which is {\it different} from the one used in
[6].
We then obtain a localization of the path integral (19) to field
configurations which are {\it different} from
the configurations that are relevant in
the  Duistermaat-Heckman integration formula: In particular, we find that our
localization provides an integration formula for the path integral (14) in
terms of equivariant characteristic classes.

If the condition (12) is satisfied, the following one-parameter family
of one-forms is in the subspace (18),
$$
\psi_{\lambda} ~=~ \frac{\lambda}{2} g_{ab} \dot\phi^{a} c^{b}
\eqno (20)
$$
The corresponding path integral
$$
Z_{\lambda} ~=~ \int [d\phi^{a}][dc^{a}]exp \{ i \int\limits_{0}^{T} \theta_{a}
\dot\phi^{a} - H + \frac{1}{2} c^{a}\omega_{ab}c^{b} + \frac{\lambda}{2}
d_{S} (g_{ab} \dot\phi^{a}c^{b}) \}
\eqno (21)
$$
is then (formally) independent of $\lambda$, and in the $\lambda\to 0$ limit
it reduces to the original path integral (14). In the following we shall be
interested in the $\lambda\to\infty$ limit of (21):

Explicitly, the action in (21) reads
$$
S ~=~ \frac{\lambda}{2} g_{ab} \dot\phi^{a} \dot\phi^{b}
+ (\theta_{a} - \frac{\lambda}{2} g_{ab}{\cal X}_{H}^{b}) \dot\phi^{a} - H
+ \frac{\lambda}{2} c^{a} ( g_{ab} \partial_{t} +
\dot\phi^{c}g_{bd}\Gamma^{d}_{ac} ) c^{b} + \frac{1}{2} c^{a} \omega_{ab} c^{b}
\eqno (22)
$$
Here $\Gamma^{d}_{ac}$ is the Christoffel symbol for the metric tensor
$g_{ab}$.
Since the path integral (21) is (formally) independent of $\lambda$, we can
evaluate it in the $\lambda \to \infty$ limit. For this, we introduce
$$
\phi^{a}(t) ~=~ \phi^{a}_{0} + \phi^{a}_{t}
\eqno (23.a)
$$
$$
c^{a}(t) ~=~ c^{a}_{0} + c^{a}_{t}
\eqno (23.b)
$$
with $\phi^{a}_{0}$, $c^{a}_{0}$ the constant modes of $\phi^{a}(t)$ and
$c^{a}(t)$, and $\phi^{a}_{t}$, $c^{a}_{t}$ the $t$-dependent
modes. We then define the path integral measure as
$$
[d\phi^{a}][dc^{a}] ~=~ d\phi^{a}_{0} dc^{a}_{0} \prod\limits_{t} d\phi^{a}_{t}
dc^{a}_{t}
\eqno (24)
$$
and introduce the change of variables
$$
\phi^{a}_{t} ~\rightarrow~ \frac{1}{\sqrt{\lambda}} \phi^{a}_{t}
\eqno (25.a)
$$
$$
c^{a}_{t} ~\rightarrow~ \frac{1}{\sqrt{\lambda}} c^{a}_{t}
\eqno (25.b)
$$
Formally, the Jacobian for this change of variables is
trivial. In the $\lambda \to
\infty$ limit we can then evaluate the path integral over the non-constant
modes, which yields
$$
Z ~=~ \int \sqrt{g} d\phi^{a}_{0}
dc^{a}_{0} { e^{ -iT H + i\frac{T}{2} c^{a}_{0} \omega_{ab}
c^{b}_{0} }  \over \sqrt{ Det [ g_{ab} \partial_{t} - \frac{1}{2}
( \Omega_{ab} + R_{abcd}c^{c}_{0} c^{d}_{0}
) ] }}
\eqno (26)
$$
Here we have defined
$$
\Omega_{ab} ~=~ \partial_{b} (g_{ac}{\cal X}^{c}_{H}) - \partial_{a} (g_{bc}
{\cal X}^{c}_{H})
\eqno (27)
$$
and $R_{abcd}c^{c}_{0} c^{d}_{0}$ is the curvature two-form. We
evaluate the determinant {\it e.g.} by the
$\zeta$-function method. This gives
$$
Z ~=~ \int \sqrt{g} d\phi^{a}_{0} dc^{a}_{0} ~ e^{ -iT H + i\frac{T}{2}
c^{a}_{0} \omega_{ab} c^{b}_{0} } \sqrt {Det\left[ { \frac{T}{2} (\Omega_{ab} +
R_{ab}) \over sinh[ \frac{T}{2} (\Omega_{ab} + R_{ab})] } \right] }
\eqno (28)
$$
Notice that if we set $H=0$, the exponential in (28) yields the Chern
character for the two-form $c^{a}_{0}\omega_{ab}c^{b}_{0}$, while
the second term in (28) gives the $\hat A$-genus for the curvature two-form
$R_{abcd}c^{c}_{0}c^{d}_{0}$. This coincides
with the result one expects from a path
integral evaluation of the Atiyah-Singer index theorem for a Dirac operator in
the background of a U(1) gauge field and a gravitational field.
Indeed, for $H=0$ the action in (28) reduces to
$$
S ~\to~ \frac{\lambda}{2} g_{ab} \dot\phi^{a} \dot \phi^{b} + \theta_{a}
\dot\phi^{a} + \frac{\lambda}{2} c^{a} (g_{ab} \partial_{t} + \dot\phi^{c}
\Gamma_{acb}) c^{b} + c^{a} \omega_{ab} c^{b}
\eqno (29)
$$
which is identical to the action used
in [7] to evaluate the Atiyah-Singer index theorem, provided we identify
$\theta_{a}$ and $\omega_{ab}$ with the
gauge field $A_{\mu}$ and its field strength
tensor $F_{\mu\nu}$.

We shall now argue, that with nontrivial $H$ the two terms in (28)
are equivariant generalizations of the Chern character (1) {\it resp.} ${\hat
A}$-genus (2) in the equivariant cohomology determined by the phase
space equivariant exterior derivative (10). Indeed, since
$$
 d_{H} (H - \frac{1}{2}
c^{a}\omega_{ab}c^{b}) ~=~ 0
\eqno (30)
$$
we can immediately identify the exponential term in (28) as an equivariant
generalization of the Chern-character (1).
Furthermore, if we introduce the covariant
generalization of the equivariant exterior derivative
$$
D_{H} ~=~ d + \Gamma + i_{H}
\eqno (31)
$$
where $\Gamma$ is the Christoffel symbol one-form, we find that
$$
D_{H} (\Omega_{ab} + R_{ab}) ~=~ 0
\eqno (32)
$$
{}From this we can conclude that the second term
in (28) is also equivariantly closed,
$$
 (d + i_{H}) \sqrt {Det\left[ { \frac{T}{2} (\Omega_{ab} + R_{ab}) \over
sinh[ \frac{T}{2} (\Omega_{ab} + R_{ab})] } \right] } ~=~ 0
\eqno (33)
$$
Hence it determines an equivariant generalization of the $\hat
A$-genus. As a consequence we have verified that formally {\it the bosonic
phase
space path integral (14) evaluates to a combination of the equivariant Chern
character and the equivariant $\hat A$-genus,} {\it i.e.} an equivariant
generalization of the Atiyah-Singer index density for a Dirac operator.

The equivariant $\hat A$-genus in (28) is determined by  expanding
$\hat A$ at $\frac{T}{2}\Omega_{ab}$ in the "variable" $\frac{T}{2} R_{ab}$.
Since the function
$$
f(z) ~=~ { \frac{z}{2} \over sinh \frac{z}{2} }
\eqno (34)
$$
has poles at $2ik\pi$, this suggests an interesting phase structure for the
pertinent thermodynamical partition function at finite temperatures $T \to
i\beta$.

We observe that our final result (28) is quite
similar to the Lefschetz formulas by
Atiyah, Bott, Singer [9] and Bismuth [10]. Indeed,
if in addition of (12) we also
assume that the hamiltonian Lie-derives the symplectic one-form,
$$
{\cal L}_{H} \theta ~=~ 0
\eqno (35)
$$
we can use loop space equivariant cohomology to write
the entire action (29) in the
form
$$
S ~=~ (d + i_{\dot\phi} + i_{H} ) ( \frac{\lambda}{2} g_{ab} \dot\phi^{a} c^{b}
+ \theta_{a} c^{a})
\eqno (36)
$$
Here we have defined
$$
i_{\dot\phi} ~=~ \dot\phi^{a} i_{a}
\eqno (37)
$$
and $i_{H}$ is now interpreted as a loop space interior multiplication.
This representation of the action admits the following interpretation:
As before, we can identify
$$
(d + i_{\dot\phi}) (\frac{\lambda}{2} g_{ab}\dot\phi^{a}c^{b} +
\theta_{a}c^{a})
{}~=~
\frac{\lambda}{2} g_{ab} \dot\phi^{a} \dot \phi^{b} + \theta_{a}
\dot\phi^{a} + \frac{\lambda}{2} c^{a} (g_{ab} \partial_{t} + \dot\phi^{c}
\Gamma_{acb}) c^{b} + c^{a} \omega_{ab} c^{b}
\eqno (38)
$$
with a Dirac operator in the background of a U(1)
gauge field $\theta_{a}$  and a gravitational field. If we denote the chiral
components of this Dirac operator by $D_{+}$
and $D_{-}$, the remaining terms in (36)
then coincide with the terms, that we expect to obtain in a supersymmetric path
integral representation of the Lefschetz number
$$
L(e^{iTH}) ~=~ Tr\{ e^{iTH} (e^{- iT D_{+}D_{-}} - e^{ - iT D_{-}D_{+} }) \}
\eqno (39)
$$
for the hamiltonian $H$. As a consequence we conclude that in the case of
hamiltonians for which the conditions (12), (35) are satisfied, our result (28)
reproduces the result by Atiyah, Bott, Singer [9] and  Bismuth [10].

{}From the  relation between (14) and (39) we can draw the following
conceptually interesting interpretation of our
result: We have found, that a bosonic
path integral (14)  can be related to the
properties of a (functional) Dirac operator
defined in the canonical phase space of the bosonic theory. In particular,
the path integral (essentially) coincides with
the Lefschetz number of the bosonic
hamiltonian evaluated for this Dirac operator. (Notice that in field theory
applications this Dirac operator is defined in an infinite dimensional space.)

In conclusion, we have investigated a general family of bosonic path integrals.
We have found that these path integrals can
be evaluated in a closed form, and the
result coincides with the equivariant generalization of the Atiyah-Singer index
density for a Dirac operator. The restriction (12) that we have introduced is
relatively mild - locally, it can be satisfied by an arbitrary hamiltonian.
This
indicates, that the corrections to our
result should in some sense reflect global
properties of the theory.  Our computations have also
been formal in that we have
overlooked {\it e.g.} difficulties associated with the definition of the path
integral measure: There might be anomalies associated with our argument that
the
path integral is  $\lambda$ independent. However, since we have also obtained a
relation between our result and the (infinitesimal) Lefschetz formula by
Atiyah,
Bott, Singer and Bismuth, we expect that in a number of interesting cases our
arguments can be made rigorous.  Finally, we conclude that our result clearly
indicates that equivariant characteristic classes can
be highly relevant to a number
of important problems in theoretical physics.

\vskip 0.4cm
A.N. thanks J.M. Bismuth and A. Reyman for discussions. Both
authors thank the Theory Division at CERN for hospitality during this work.
A.N.
also thanks  Ecole Polytechnique and Paris VI for hospitality during this work.

\vfill\eject

{\bf References}

\vskip 0.8cm

\begin{enumerate}

\item For a review, see T. Eguchi, P.B. Gilkey and A.J. Hanson,
Phys. Repts. {\bf 66} (1980) 213

\item E. Witten, Comm. Math. Phys. {\bf 117} (1988) 353; For a review, see D.
Birmingham, M. Blau, M. Rakowski and G. Thompson, Phys. Repts. {\bf 209} (1991)
129

\item E. Witten, J. Diff. Geom. {\bf 17} (1982) 661

\item N. Berline and M. Vergne, Duke Math. Journ. {\bf 50} (1983) 539

\item For a review, see N. Berline, E. Getzler and M. Vergne, {\it Heat Kernels
and Dirac Operators} (Springer-Verlag, Berlin, 1991)

\item M. Blau, E. Keski-Vakkuri and A.J. Niemi, Phys. Lett. {\bf B246} (1990)
92; A.J. Niemi and P. Pasanen, Phys. Lett. {\bf B253} (1991) 349; E.
Keski-Vakkuri, A.J. Niemi, G. Semenoff and O. Tirkkonen,
Phys. Rev. {\bf D44} (1991) 3899

\item A. Hietam\"aki, A.Yu. Morozov, A.J. Niemi and K. Palo, Phys. Lett. {\bf
B263} (1991) 417;  A. Yu. Morozov, A.J. Niemi and K. Palo, Physics Letters {\bf
B271} (1991) 365; Nucl. Phys. {\bf B} (to appear)

\item J.J. Duistermaat and G.J. Heckman, Inv. Math. {\bf 72} (1983) 153;
N. Berline and M. Vergne, Amer. J. Math, {\bf 107} (1985) 1159

\item M. Atiyah and R. Bott, Ann. Math. {\bf 86} (1967) 374; {\it ibid.} {\bf
88}, (1968) 451; M. Atiyah and I. Singer, Ann. Math. {\bf 87} (1968) 546; {\it
ibid.}  {\bf 93} (1971) 119

\item  J.-M. Bismut, Comm. Math. Phys. {\bf 98} (1985) 213; {\it ibid.} {\bf
103} (1986) 127; J. Funct. Anal. {\bf 62} (1985) 435

\end{enumerate}

\end{document}